# A High-Speed Waveguide Integrated InSe Photodetector on SiN Photonics for NIR Applications

Srinvasa Reddy Tamalampudi[1], Juan Esteban Villegas[1,2], Ghada Dushaq[1], Raman Sankar[3], Bruna Paredes[1] and Mahmoud Rasras[1*]

[1] Department of Electrical Engineering, NYU Abu Dhabi, Saadiyat Island, Abu Dhabi, AD, UAE.
[2] Department of Electrical Engineering, NYU Tandon School of Engineering, 6 MetroTech Center, Brooklyn, 11201, NY, USA.
[3] Institute of Physics, Academia Sinica, Nangang, Taipei, Taiwan
S.R.T. and J.E.V. contributed equally
*Corresponding author: mr5098@nyu.edu;

**Abstract**

On-chip integration of two-dimensional (2D) materials offers great potential for the realization of novel optoelectronic devices in different photonic platforms. In particular, indium selenide (InSe) is a very promising 2D material due to its ultra-high carrier mobility and outstanding photo-responsivity. Here, we report a high-speed photodetector based on a multilayer 90 nm thick InSe integrated on a silicon nitride (SiN) waveguide. The device exhibits a low dark current of ~10 nA at 1V bias, a remarkable photoresponsivity of 0.38 A/W, and high external quantum efficiency of ~ 48.4% measured at 5 V bias. This performance is tested at near-infrared (NIR) 976 nm wavelength under ambient conditions. Furthermore, using numerical and experimental investigations, the estimated absorption coefficient per unit length is 0.11dB/µm. To determine the dynamic response of the photodetector, its small and large signal frequency response are also evaluated. A 3-dB radiofrequency (RF) bandwidth of 85 MHz is measured with an open-eye diagram observed at 1 Gbit·s$^{-1}$ data transmission. Given these outstanding optoelectronic merits, active photonic devices based on integrated multilayer InSe can be realized for a variety of applications including short-reach optical interconnects, LiDAR imaging, and biosensing.

## 1. Introduction

The development of near-infrared (NIR 750 nm-1064 nm) integrated photodetectors is crucial for a wide range of modern applications spanning lab-on-a-chip devices and environmental sensing to short-range communication in data centers[1,2]. Additionally, with the progress of autonomous vehicle technologies, LiDAR-based sensors are often made between 940-980 nm wavelengths for short- to medium-range positioning and mapping[3,4]. The absorption of silicon in this bandwidth requires the use of an alternative platform to build integrated photonic devices. In this instance, silicon nitride (SiN) can be used to construct optical circuits, while Si, Ge, and III-V semiconductors to fabricate active optical components such as photodetectors [5-7]. The SiN-based waveguides are fully compatible with complementary metal-oxide semiconductor (CMOS) processing and offer low optical propagation losses, high integration densities, and a broadband optical transparency window from the visible to the infrared[8,9]. In addition, they exhibit a linear transmission even at high input optical powers[10]. The SiN waveguide NIR photodetectors have been demonstrated with standard semiconductor platforms while silicon (Si)-based photodetectors have shown a higher responsivity than Ge/III-V at 850-1000 nm wavelength[2]. However, their bandwidths are limited due to the high RC time constant[11-13]. Therefore, it is important to identify novel active photonics materials that can be integrated on the SiN platform to realize NIR high-speed components, including optical modulators and photodetectors.

In the past decade, photodetectors based on graphene and several other two-dimensional materials have demonstrated exceptionally high responsivity and quantum efficiency from visible to near-infrared wavelengths[14-19]. Inherently, these materials stack together out of the plane by weak Van der Waals (VdW) forces, and they are covalently bonded in the in-plane orientation. They can be easily integrated into various platforms, simplifying the fabrication process. Consequently, on-chip integrated 2D photodetectors have gained a lot of attention due to their advantage compared to traditional bulk semiconductors photodetectors[20, 21]. Particularly, graphene photodetectors with bandwidths exceeding 100 GHz on the silicon photonics platform have been demonstrated[22]. However, most of these photodetectors have been studied for telecom bands[20, 23, 24], while integrated NIR photodetectors are not yet well reported.

Recently, layered Indium Selenide (InSe) has gained renewed interest as an active optical material due to its anisotropic optical, electrical, and mechanical properties. It has been proposed for several applications including memory devices, optical sensors, artificial synapses, and neural network computing[25, 26]. This 2D material has a thickness-dependent tunable bandgap and it belongs to the group IIIA-VIA layered semiconductors (MX, M = Ga and In, X = S, Se, and Te)[27]. Recent studies have reported that field-effect devices based on atomically thin InSe flakes present unique properties such as a high carrier mobility of $10^4$ $cm^2$ $(Vs)^{-1}$ and $10^3$ $cm^2$ $(Vs)^{-1}$ at liquid nitrogen and room temperatures, respectively[26, 28]. The carrier mobility of the InSe nanosheet is much larger than that of transition metal dichalcogenides (TMDs) and is analogous to that of black phosphorus[29]. Large-scale growth of InSe thin films on various substrates has been realized using chemical vapor deposition(CVD)[30], atomic layer deposition (ALD)[31], and pulsed laser deposition (PLD)[32] methods, demonstrating the feasibility of large-scale integration of InSe devices.

Photodetectors based on few-layered InSe sheets exhibit a broadband response from visible to near-infrared wavelengths with high photoresponsivity[33]. Furthermore, extensive reports have demonstrated different types of InSe detectors based on photoconductive (PC), phototransistors (PT)[33], self-powered photodetectors (SPPD)[34, 35], and avalanche photodetectors (APD)[36]. However, most of these studies are limited to vertically coupled free space devices in the visible range, where the InSe has the highest absorption coefficient. Recently, Xiaoqi Cui et. al., demonstrated an on-chip integrated InSe photodetector operating at 520 nm wavelength with a responsivity of 0.115 A/W but the study is limited to the static response[37]. Despite its promising performance, the scarcity of InSe studies in NIR for integrated systems inspires an in-depth investigation of this material on the SiN platform.

In this work, we demonstrate integrated InSe photodetectors on the SiN platform operating in the NIR region. The hybrid SiN/InSe waveguide photodetectors are fabricated by mechanical exfoliation of InSe flakes on top of SiN waveguides. The device operates under bias and exhibits a low dark current of a few nano-amps. Furthermore, a remarkable photoresponsivity of 0.38 A·W$^{-1}$ and low noise-equivalent power (NEP) of 4.7 nW·Hz$^{1/2}$ are achieved at 976 nm wavelength. The devices exhibit a 3-dB radiofrequency (RF) bandwidth of 85 MHz and open-eye diagrams are demonstrated for on-off modulated signals from 50 Mbit·s$^{-1}$ to 1 Gbit·s$^{-1}$. The achieved excellent merits offer a new avenue for building and designing active photonic components on the SiN platform using InSe.

## 2. Results and Discussion

### 2.1 InSe integration into SiN photonics

Figure 1a shows an optical microscopy image of the fabricated unloaded SiN waveguide chip. In this design, a 400 nm top silicon nitride layer and a 2 µm buried oxide are used. The waveguide is air-cladded and has a width of 1.1 µm. Light is coupled in and off the SiN chip using lensed fibers. The photodetector's contacts are formed using a pre-patterned metal electrode of Cr/Au (10 nm/120 nm). Figure 1b shows a schematic representation of the 3D cross-section view and a schematic representation of the photodetector, respectively. Note that the metal pads and the SiN waveguide are not on the same plane and there is a 270 nm height difference between the metal and waveguide top surfaces.

A deterministic dry transfer procedure is used to transfer the mechanically exfoliated multilayered InSe on top of the SiN waveguide, laid directly on top of the metal electrodes. It is worth noting that a recent study has reported an anisotropic nature of InSe where the optical absorption is highly affected by the crystal orientation. Given our method for transfer, the fabricated devices are not perfectly aligned with the crystal orientation that maximizes the optical absorption[34]. More details on the deterministic transfer process have been reported in our previous works[23, 38]. A scanning electron microscopy (SEM) image of a fabricated device is shown in Fig. 1(c). The metal electrodes are symmetrically deposited on the sides of the waveguide spaced with a gap fixed at 8 µm (distance between metal edges), which minimizes optical absorption losses from metal absorption in the electrodes. In the fabricated device, the overlapping length of the detector is ~30µm along the waveguide. The thickness of the InSe is measured using an atomic force microscope which is determined to be ~ 90nm (see Fig. 1d). As can be seen in this figure, the flake adheres conformally to the photonic waveguide structure beneath it.

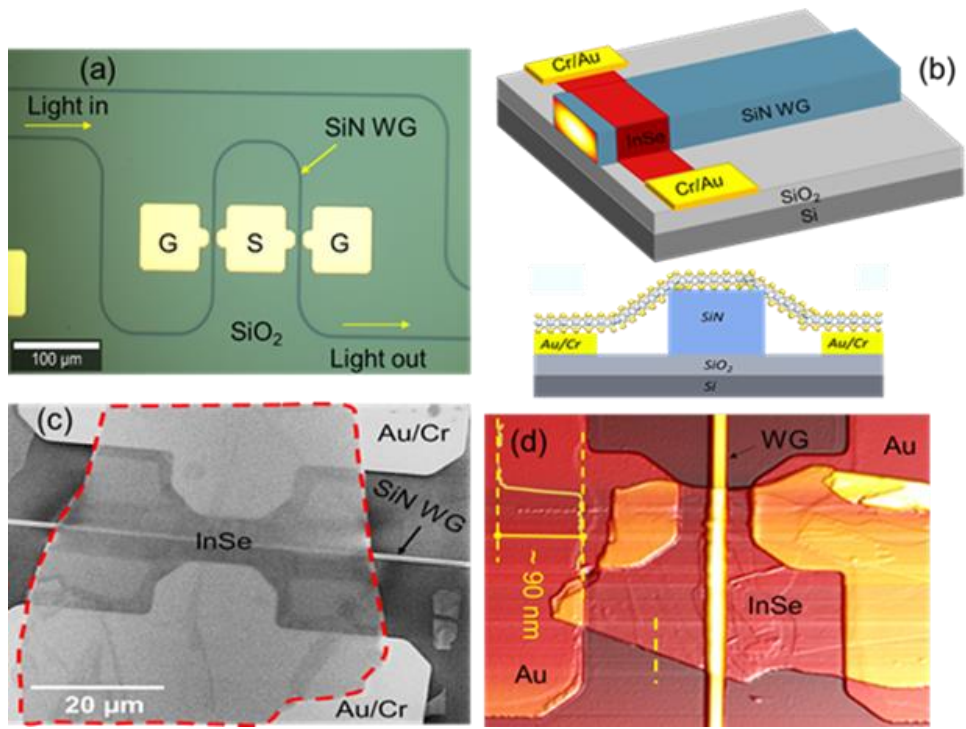

**Figure 1. Hybrid integration of InSe into SiN photonic circuit** (a) optical image of the SiN unloaded waveguide chip. (b) 3D cross-section and schematic representation of heterogeneous InSe/SiN photodetector. (c) A SEM image of the transferred InSe on SiN waveguide. (f) An AFM image scan, the dashed yellow line shows a thickness of ~ 90 nm InSe and interaction length of ~ 30 µm.

## 2.2 Guided Light Interaction with Multilayer InSe

The material characteristics of bulk and transferred InSe are confirmed by energy dispersive spectroscopy (EDS), Raman spectroscopy, x-ray diffraction spectroscopy (XRD), and ellipsometry. To confirm its stoichiometry, EDS of the grown single crystal InSe is performed on the selected area depicted by the highlighted pink color box shown in the SEM image in Fig. 2a. The spectra indicate only peaks for the indium (In) and selenium (Se) elements with an atomic ratio about 1:1 (In:Se). Additionally, the SEM image and elemental composition of the waveguide integrated InSe are shown in Fig. S1, confirming a controlled 1:1 (In:Se) stoichiometry after exfoliation and transfer. To confirm the crystallographic structure and the material phase, room temperature XRD on the bulk crystal is carried out. Results are shown in Fig. 2(b). All the observed diffraction peaks agree well with the previously reported patterns of β-phase InSe, confirming the high purity of the sample[33, 39]. Figure S2 shows the schematic representation of the β-phase InSe crystal structure. Furthermore, Raman spectra of the waveguide-integrated InSe also confirm the β-InSe phase as depicted in the supplementary Fig. S3. All observed active phonon modes are in agreement with other reports on β-phase InSe[40]. As shown in Fig. S4, high-resolution spectroscopic ellipsometry performed on InSe flakes on 300 $SiO_2$/Si substrate is used to determine its refractive index (n) and extinction coefficient (k). This data is fed into optical simulations.

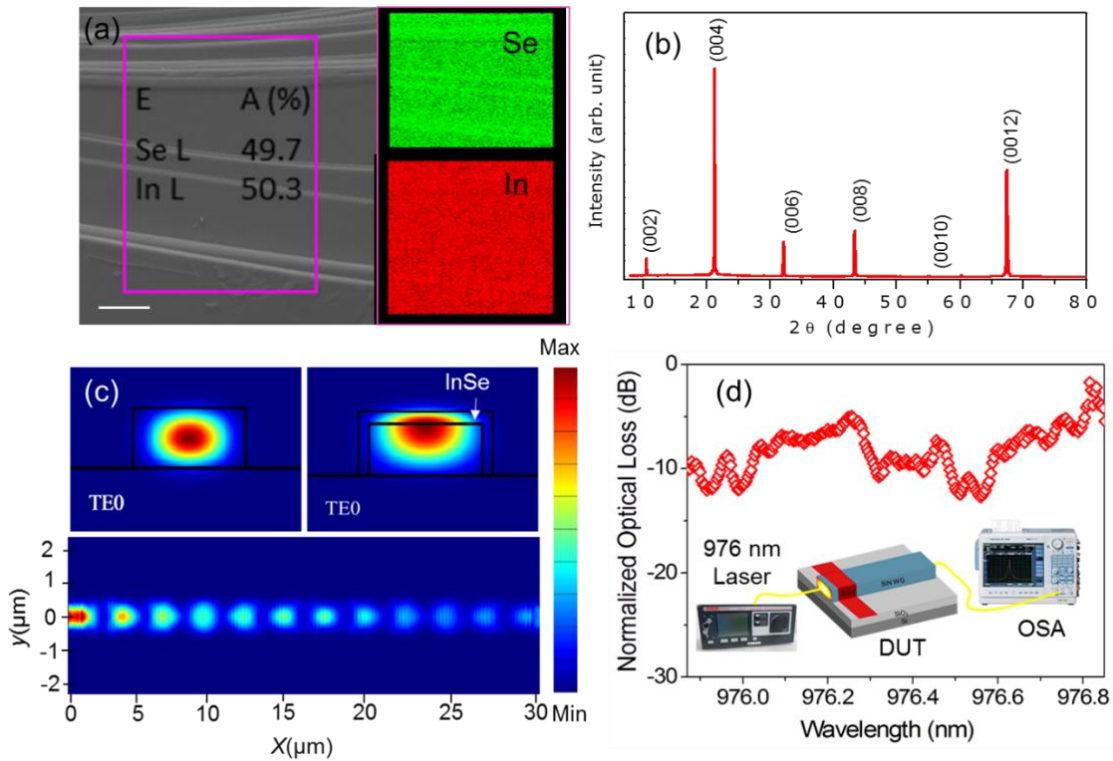

**Figure 2. Material characterization and guided light interaction** of 90 nm thick InSe layer with a length of 30 µm at 976 nm wavelength (a) SEM scan with EDS maps captured from the multilayer InSe. The scale bar is 50 µm. (b) XRD spectrum of the bulk β-InSe. (c) electric-field profiles ($|E|^2$) of TE modes of unloaded SiN waveguide and 90 nm InSe on SiN at 976 nm (top panel). FDTD simulation of beam propagation (bottom panel) (d) measured optical propagation losses of InSe. Inset shows a schematic representation of the measurement setup.

To study the device's optical characteristics, numerical investigations of the hybrid InSe/SiN structure's effective refractive index ($n_{eff}$), mode confinement, and light absorption loss ($\kappa$) are carried out for a 90 nm InSe thick flake. The modal and propagation properties are simulated using an eigenmode solver and finite differences in the time domain (FDTD) (see methods). Figure 2(c) (top panel) shows the electric-field profiles ($|E|^2$) of the fundamental transverse electric field modes ($TE_0$) of the unloaded SiN waveguide and loaded structures. Supplementary table S1 encloses simulated information about the waveguide parameters. Since the InSe flake has a higher refractive index (n= 3.1) than the SiN waveguide (n= 2.7); the optical mode is more confined to the InSe flake, leading to higher optical absorption in that layer.

The evanescent field coupling in the hybrid structure over a propagation length is simulated using FDTD as shown in Fig. 2c (bottom panel). The calculated $TE_0$ mode optical propagation losses for the loaded SiN waveguide is 0.27 dB/µm. It is evident that an interaction length of 30 µm and a flake thickness of 90 nm results in near-complete absorption, offering a reduced device footprint. Furthermore, the optical losses before and after the integration of multilayer InSe are measured by coupling the laser light into a reference waveguide and monitoring the output power using an optical spectrum analyzer (OSA). Figure 2d depicts the introduced losses, calculated from measured transmissions of loaded and unloaded waveguides (the inset shows a schematic of the setup). This result is based on an InSe thickness of 90 nm and an interaction length of ~ 93 µm, at 976 nm. The measured optical loss due to the absorption in the loaded device is 0.11 dB/µm, which is in the same range as the simulated losses. The discrepancy between the calculated and measured propagation losses can be attributed to variability and the dependence on the absorption of the InSe crystal's orientation, as discussed earlier.

## 2.3 Static Electrical and Optical Response

The current-voltage characteristics of the integrated photodetector are tested in the dark state and under monochromatic laser excitation of 976 nm. As previously stated in section 2.2, the refractive index contrast between InSe and SiN induces high optical confinement within the active InSe layer (see Fig. 2c). In this case, InSe absorbs photons via direct band-to-band transitions, which is confirmed by the measured photoluminescence of InSe (see Fig. S5. It also leads to an increase in electrical conductivity due to the electron-hole pairs generation. Additionally, the photo-excited carriers drift using the lateral electric field applied between the two in-plane metal electrodes.

Figure 3(a) shows the I–V characteristics of the photodetector under dark and at different coupled light intensities, in a range between 0.21 µW and 340 µW. As shown in the inset of this figure, the devices exhibit a low dark current ~ 10 nA at 1V and increases to about ~ 94 nA at 5 V bias (device area ~66.5 µm$^2$). Note that due to the anisotropic transport behavior of InSe, the dark current values also depend on the crystal orientation. This effect has been studied by Z. Guo et al., demonstrating an angle-dependent dark current in InSe with a ratio of 3.76 ($V_{ds}$ = 1V ) between two orthogonal orientations[40]. At a coupled light intensity of 0.21 µW, an open-circuit voltage ($V_{oc}$) and short-circuit current ($I_{sc}$) of 0.19 V and 63 pA, are measured, respectively. Increasing the light intensity further to 340 µW leads to a $V_{oc}$ and $I_{sc}$ of 0.61 V and 73 nA, respectively.

It is worth noting that the work function difference between Au (5.1 eV) and InSe (4.6) results in a Schottky junction[41, 42]. Hence, the fabricated device is comprised of back-to-back Schottky diodes. However, it is clear that the device exhibits an asymmetric transport behavior. This indicates that the Schottky barrier height at the two Au-InSe junctions is not the same. This can be attributed to defects and trap states at the interface between metal and semiconductor, which are difficult to control and can result in fermi-level pinning[43]. Additionally, the asymmetric contact geometries, defined as the difference in the contact area between the InSe flake and the metals on both sides, may result in a rectifying behavior despite the symmetric metal electrode configuration[44]. This has been reported previously in InSe and other 2D materials, where a net photocurrent at zero bias is observed enabling a self-powered photodetector[34, 35].

The generated photocurrents ($I_{ph} = I_{light} - I_{dark}$) under different powers as a function of the applied voltage are depicted in Fig. 3b. It shows a high photogenerated current which can be attributed to the InSe's direct bandgap nature and the extended interaction length with the optical guided mode. The photocurrent *vs* incident optical power curve (see Fig. S6a) indicates a nonlinear behavior for small optical powers, which may be primarily attributed to the presence of long-lived traps in the InSe[45].

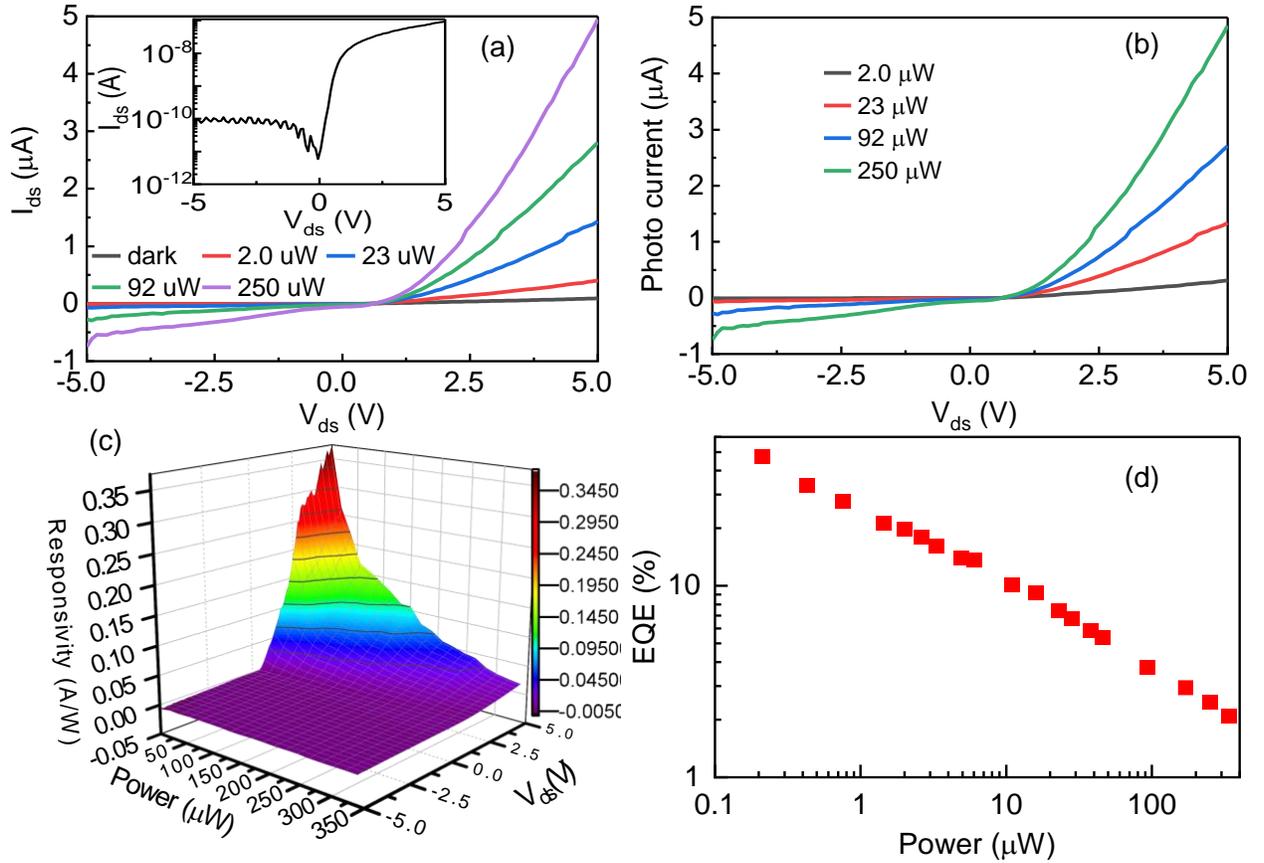

**Figure 3. Static response of InSe photodetector** (a) I-V characteristics of the waveguide integrated InSe photodetector under dark and different powers for 976 nm wavelength, respectively. Inset shows the dark current of the device. (b) The generated photocurrent for different coupled laser powers. (c-d) The measured photocurrent, responsivity, and EQE are at different powers.

The performance of the photodetectors is also assessed by measuring their responsivity ($R = I_{ph}/P$). It is calculated as the ratio of the photocurrent ($I_{ph}$) to the optical power (P) received by the detector. Figure 3c shows a 3D plot of photocurrent and responsivity as a function of applied voltage. A photoresponsivity of 0.38 AW$^{-1}$ at 5 V is measured. This corresponds to an external quantum efficiency (EQE = 1.24 × R/λ) of 44%. Based on the obtained results, the photocurrent increases with the applied voltage, unlike the bolometric effect where the photocurrent has an opposite sign to the applied bias. Additionally, since the detector has a symmetric electrode configuration, hence, the photo-thermoelectrical effect (PTE) can be ruled out. Therefore, our photodetector's performance is dominated by the photoconductive effect.

Furthermore, the photodetector's noise equivalent power is calculated to assess its minimal detectable power (noise equivalent power NEP). The latter is defined as $i_n/R$, where $i_n$ is the noise current and R is the responsivity. Therefore, the NEP indicates the excitation power needed to produce a photocurrent equal to the noise current. There are typically three noise sources present: $1/f$ noise, shot noise, and Johnson noise, where $f$ is the frequency. However, for high-frequency signals $f > 1$ kHz, the total noise current is determined by the Johnson noise ($i_{nJ} = (4k_B T \Delta f)/R_0$, where $k_B$ is the Boltzmann constant, T is the temperature, $R_0$ is the channel resistance) and shot noise ($i_{ns} = 2e(I_d + I_{ph})\Delta f$, where q is the electric charge, $I_D$ is the dark current, $\Delta f$ is the device bandwidth)[46, 47]. The measured Johnson noise, shot noise, and NEP at 5 V bias are 0.17 nA/Hz$^{1/2}$, 1.68 nA/Hz$^{1/2}$, and 4.7 nW·Hz$^{1/2}$, respectively (See Fig. S6b).

## 2.4 High-speed Photo Response

To test the dynamic performance of the photodetector, its frequency response (RF) is measured. A continuous wave laser at 976 nm is modulated by a commercial optical high-speed modulator. A DC source is used to bias the photodetector using a bias tee. The RF signal is then fed into the network spectrum analyzer (ESA). The schematic of the experimental setup is shown in supplementary Fig. S7.

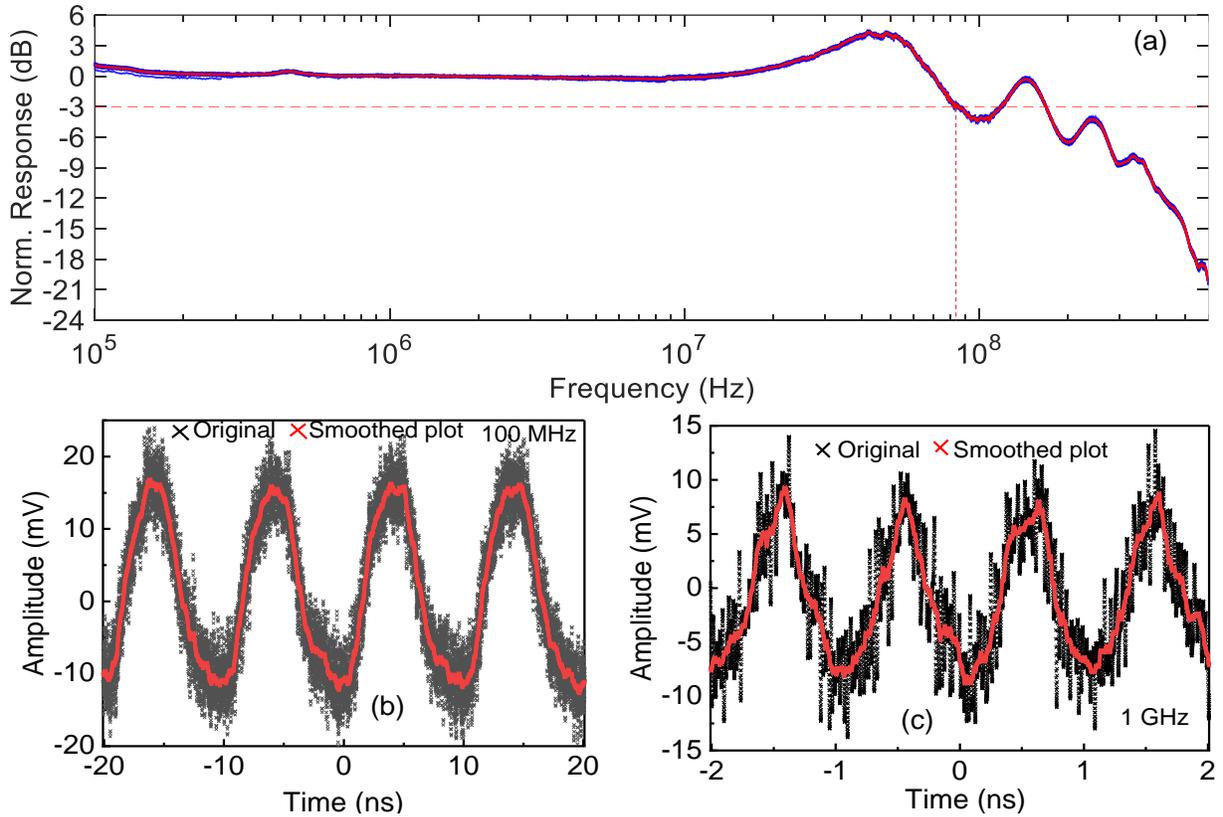

**Figure 4. Dynamic performance characteristics of the InSe photodetector** (a) Normalized frequency response at 10V. A total of 50 measurements (blue scattered points) and average (red line) data are plotted. A 3-dB cutoff frequency of 85 MHz is measured. (b-c) the temporal response recorded in the time domain for 100 MHz and 1GHz modulated optical signals, the curves are smoothed by using 200 points moving average.

Figure 4a shows the average frequency response of the InSe photodetector. The device exhibits a robust reproducible response with a 3-dB cutoff frequency of 85 MHz at 10V bias. The extracted rise time $\tau_{rise} \approx 0.35/f_{3dB}$ is approximately 4.1 ns, where $\tau_{rise}$ is the time between the 10% and 90% transient response, equivalent to a time constant of τ = 1.87 ns. To the best of our knowledge, this InSe-based photodetector has the smallest recorded response time, several orders of magnitude smaller than earlier investigations at the NIR band, and particularly 5 to 6 orders of magnitude smaller than other InSe studies (see below Table 1).

| Material | λ (nm) | R (A/W) | Dark current (nA) | Response time | Ref. |
|---|---|---|---|---|---|
| Doped InSe | 980 | $7.87 * 10^3$ at 1V | 200 | 5 ms | [48] |
| Gr/GaAs | 980 | 0.059 at 0 V | 100 | 24 ms | [49] |
| Te/Ge | 980 | 0.5 at 0 V | 50 | 14 ms | [50] |
| p-n-InSe | 980 | 0.0005 at 0V | 0.0015 | 8.3 ms | [51] |
| InSe | 980 | 0.00015 at 2 V | | 18 ms | [51] |
| InSe | 976 | 0.38 at 5 V | 97 | 4.1 ns | This work |

*Table 1. Comparison of the Performance of InSe/SiN Photodetector with Other NIR Photodetectors*

The temporal response of an InSe device is studied using an arbitrary waveform generator and high-speed optical modulator, and the output signal shape is monitored using a high-speed oscilloscope. The complete setup is shown in the supplementary Fig. S8 and the voltage-dependent $f_{3dB}$ is depicted in Fig. S9. The measured and smoothed responses at $V_{ds}$ = 10 V are shown in Fig. 4 (b and c) using a modulated sinusoidal signal. The signal to noise ratio of the temporal responses are evaluated to be 5.8 and 3.4 at 100 MHz and 1 GHz, respectively. It's important to note that the signal can be clearly discerned at 1 GHz. The full width half maximum (FWHM) values of the temporal responses are measured to be 4.3 ns and 0.41 ns at 100 MHz and 1 GHz, respectively. The photodetector is further tested for an on-off modulated pseudo-random binary sequence (PRBS) data transmission. Figure 5 shows the measured eye diagrams of the photodetector at 10 V bias. A clear and open eye diagram is obtained for NRZ (non-return-to-zero) data rates 0.1 Gbit·s$^{-1}$. Additionally, an open eye diagram is recorded at a bit rate of 0.5 and partially open one at 1 Gbit.s$^{-1}$. The performance of the SiN/InSe photodetector is compared to other waveguide-integrated photodetectors in Fig. 6[52-58]. Notably, most integrated photodetectors are reported for telecom bands, and this study's $f_{3dB}$ is lower compared to graphene or black-phosphorous (BP)-based photodetectors. From the capacitance-resistance (RC) measurements, the measured values are ~ 0.1 pF, and 0.1MΩ at 1 MHz frequency, respectively, which corresponds to RC limited bandwidth of ~100 MHz

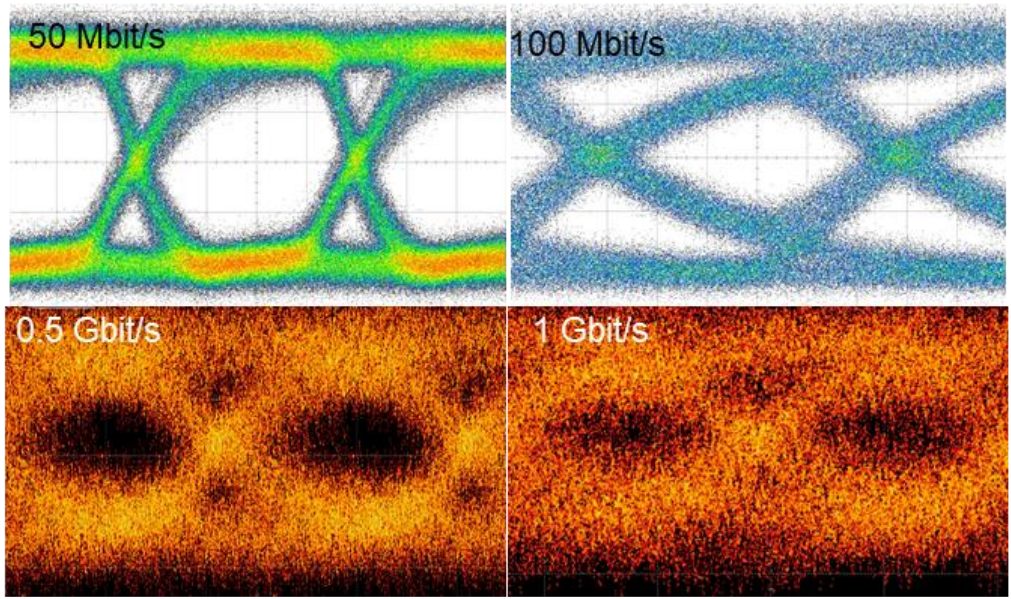

**Figure 5. Measured eye diagram** at data rate from 50 Mbit/s to 1 Gbit/s NRZ modulation at 10 V bias.

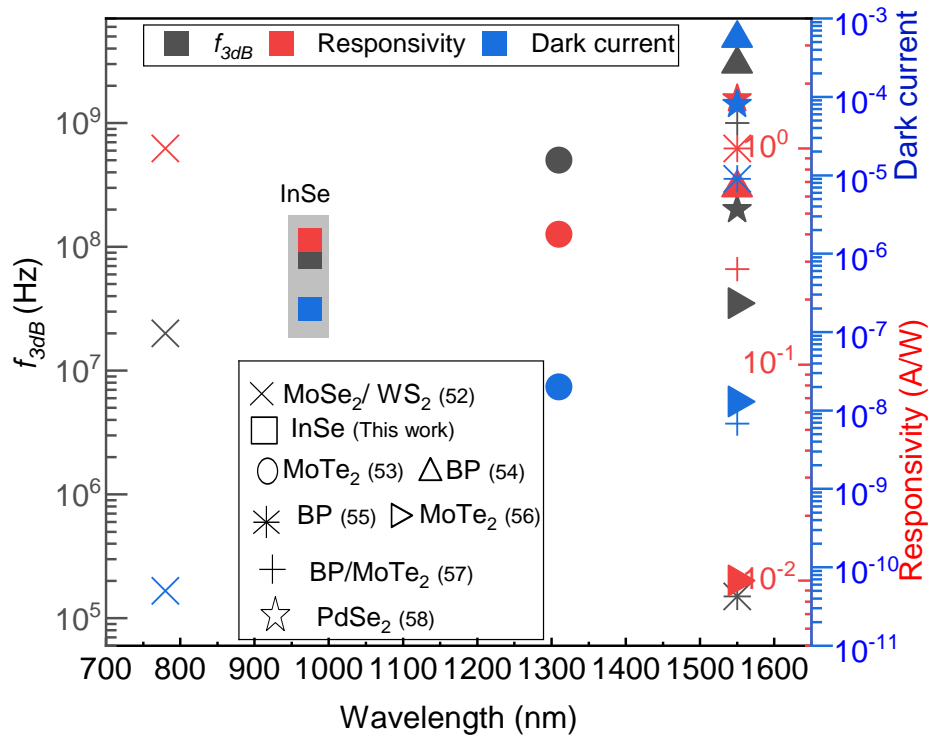

Figure 6: **A performance summary of the integrated photodetector** based on 2D materials which are previously reported showing their operating spectral response, bandwidth, and responsivity along with the InSe photodetector presented in this work.

It is evident that there is room for enhancing the bandwidth of the InSe device. One can improve the performance of the integrated InSe device by the following methods: (i) Shorten the InSe channel length to further increase the bandwidth. Photodetectors with a channel length of ~ 10 μm, need mobility of ~1000 cm$^2$/V. s to reach GHz range bandwidth (ii) In the current study, InSe flake is transferred onto pre-deposited metal electrodes, which results in a higher contact resistance compared to the buried devices. (iii) InSe exhibits a polarization-dependent anisotropic ratio of dark current to photocurrent, therefore, a control over the InSe flake orientation during transfer can enhance the device performance. (iv) Creating a heterojunction with graphene or using graphene as electrodes. (v) Encapsulation of the InSe with hexagonal boron nitride (hBN), can yield higher carrier mobility. Note that no passivation before the transfer of InSe on the SiN waveguide is used. It is well established that a transfer of hBN before the integration can passivate the waveguide and reduce roughness.

## 3. Conclusion

In summary, we demonstrated an InSe/SiN waveguide integrated photodetector operating in the near-infrared (976 nm) regime. The fabrication process is simple compared to other integrated photodetectors. The as-assembled devices exhibited an extrinsic responsivity of 0.38 A/W (EQE of ~ 44%), a low dark current of 1 nA at 1V, and a fairly low NEP of 4.7 nW/Hz$^{1/2}$. Additionally, a high-index contrast of InSe compared to SiN waveguide, (Δn) of ~ 0.4, it results in improved optical absorption. The measured absorption coefficient for a 90 nm thick flake is 0.11 dB/μm. Additionally, the device showed a 3dB cutoff frequency of 85 MHz at 10 V and open eye diagrams at a bit rate of 0.5 up to 1 Gbit.s$^{-1}$. This InSe-based photodetector exhibits high bandwidth and outperforming similar reported devices by more than five to six orders of magnitude. We believe that the bandwidth can be further improved by increasing the carrier mobility, enhancing the waveguide-InSe interface, and reducing the circuit capacitance. This study provides a realization of high-performance photodetectors using a simple architecture with potential applications in a wide range of functions including lab-on-a-chip devices, LiDAR system imaging, and short-reach interconnects for data centers.

## 4. Methods

### 4.1 Materials and fabrication

*Crystal growth*

The InSe single crystals are produced using a 99.999% pure molar combination of In (52.4% weight percentage) and Se (47.6%) compounds that are purchased from Sigma-Aldrich. Conical quartz ampoules evacuated to 10$^{-4}$ Pa are used to produce single-crystalline InSe flakes. A horizontal furnace is used to homogenize the batches and create the InSe flakes for 48 hours at 550°C. The Bridgman vertical method is used to grow the mixed crystals. The melt-filled ampoules are heated at 850°C for 24 hours before pulling; once the melt had filled the ampoule's tip, the ampoules are lowered via a 1°C temperature gradient at a rate of 0.1mm/h. The InSe crystals that are formed have dimensions of 3 cm and 1.2 cm.

*Si$_3$N$_4$ platform*

The passive SiN photonic devices are fabricated with standard 220-nm-SOI processes in Applied nano tools inc., Metal pads in the devices are patterned by electron-beam lithography followed by the deposition of metal pads (Ti/Au, 10/100 nm) by electron-beam evaporation and a standard lift-off process. After depositing the metal pads, InSe is transferred with the help of a PDMS film.

*InSe transfer*

A few layers InSe flakes are exfoliated using scotch tape and transferred onto the PDMS substrate. By using the controllable dry transfer method, the selected InSe flakes are transferred onto the SiN waveguides.

### 4.2 Simulations

The electric field profile in the silicon nitride waveguide and the beam propagation are calculated using the MODE Solutions eigenmode solver and FDTD simulation in Lumerical. The optical parameters used for InSe are those extracted from ellipsometry.

### 4.3 Material characterization

*Spectroscopic imaging ellipsometer*

The optical parameters of multilayer InSe are determined by Accurion's Imaging Ellipsometry (https://accurion.com/company). This system combines optical microscopy and ellipsometry for spatially resolved layer-thickness and refractive index measurements. The tool is highly sensitive to ultrathin single- and multi-layer films, ranging from mono-atomic or monomolecular layers (sub-nm regime) up to thicknesses of several microns. Additionally, Imaging Ellipsometers can perform layer thickness measurements with a spatial resolution down to 1 µm. The ellipsometric parameters (Psi ($\psi$) and Delta ($\Delta$)) are fitted using EP4 model software.

*Scanning electron microscopy*

A (FEI) Quanta 450 field emission scanning electron microscope with an electron energy of 10 KV is used to image the photonic chips while they are placed on an SEM stub using carbon tape.

*Atomic force microscopy*

The tapping mode of the WITec Atomic Force Microscope (AFM) module is used to determine the thickness of the transferred InSe flake. The cantilever tip (Scanasyst-air) had a radius of 7 nm, a force constant of 0.2 N/m, and a resonance frequency of 14 kHz.

### 4.4 Device characterization

*Optical characterization*

The optical transmission is performed using edge coupling the light into the photonic chips through lensed fiber and a tunable laser operating at 976 nm. The output response from the devices is collected by an output lensed fiber and detected by a power meter. The output optical power intensities are calibrated before testing the device using a standard photodiode power sensor.

*DC measurements*

The steady-state performance of the InSe photodetectors is tested by measuring their dark current and responsivity. A transverse electric-polarized (TE) of 976 nm light is edge coupled via lensed optical fiber to the devices. A curve tracer/power device analyzer / (Agilent B1505A) is used to control the biases and measure the I-V characteristics in the dark and upon light coupling via a pair of standard DC electrical probes.

## Acknowledgments

NYUAD Research Enhancement Fund supported this work. Part of the research work was carried at the Core Technologies Platform (CTP) at NYU Abu Dhabi. The authors are thankful to the CTP staff for their technical and instrumentation support, especially, Mr. Nikolas Giakoumidis, Dr. James Weston and Dr. Qiang Zhang for the optical, analytic and microfabrication support. RS acknowledge the financial support provided by the Ministry of Science and Technology in Taiwan under project numbers NSC-111-2124-M-001-009 & AS-iMATE-111-12.

## Contribution

S.R.T. initiated the study and planned the experiments. S.R.T. fabricated and measured the electrical and photodetector performance of the devices. J.E.V. and S.R.T. performed RF and bandwidth measurements and analyzed the data. S.R. T. performed the material characterization (EDS, Raman, XRD and ellipsometer) of InSe crystal and integrated devices. J.E.V. performed the FDTD simulations. J.E.V. and G.D performed the SEM and AFM characterization. R. S. performed the growth of the InSe crystals. B.P., G.D., and J.E.V., design the SiN photonic chips. All the authors analyzed and discussed the data. S.R.T., J.E.V., G.D., and M. R. wrote the paper. S.R.T. and J.E.V. contributed equally. M.R supervised the project.

## Disclosures

The authors declare no competing interests.

## Data availability

Data underlying the results presented in this paper are not publicly available at this time but may be obtained from the authors upon reasonable request.

## Supplemental document
Supplementary Figs. S1–S9, Tables S1.